\documentclass[english,aps,manuscript]{revtex4}
\usepackage[T1]{fontenc}
\usepackage[latin9]{inputenc}
\usepackage[active]{srcltx}
\usepackage{float}
\usepackage{units}
\usepackage{amsmath}
\usepackage{graphicx}

\makeatletter
\@ifundefined{textcolor}{}
{%
 \definecolor{BLACK}{gray}{0}
 \definecolor{WHITE}{gray}{1}
 \definecolor{RED}{rgb}{1,0,0}
 \definecolor{GREEN}{rgb}{0,1,0}
 \definecolor{BLUE}{rgb}{0,0,1}
 \definecolor{CYAN}{cmyk}{1,0,0,0}
 \definecolor{MAGENTA}{cmyk}{0,1,0,0}
 \definecolor{YELLOW}{cmyk}{0,0,1,0}
}

\makeatother

\usepackage{babel}
\begin{document}

\title{The effect of the size of the system, aspect ratio and impurities
concentration on the dynamic of emergent magnetic monopoles in artificial
spin ice systems.}

\author{Alejandro León}

\affiliation{Facultad de Ingeniería, Universidad Diego Portales, Santiago Chile}
\begin{abstract}
In this work we study the dynamical properties of a finite array of
nanomagnets in artificial kagome spin ice at room temperature. The
dynamic response of the array of nanomagnets is studied by implementing
a \textquotedblleft{}frustrated celular autómata\textquotedblright{}
(FCA), based in the charge model and dipolar model. The FCA simulations,
allow us to study in real-time and deterministic way, the dynamic
of the system, with minimal computational resource. The update function
is defined according to the coordination number of vertices in the
system. Our results show that for a set geometric parameters of the
array of nanomagnets, the system exhibits high density of Dirac strings
and high density emergent magnetic monopoles. A study of the effect
of disorder in the arrangement of nanomagnets is incorporated in this
work.
\end{abstract}
\email{alejandro.leon@udp.cl}


\maketitle

\section{Introduction}

Based on Dirac\textquoteright{}s prediction about the existence of
magnetic monopoles, considering the quantization of the electrical
charge {[}1{]}, many research teams have attempted to find these monopoles
{[}2{]}, but their results have been inconclusive. Attention is currently
directed at the study of elemental excitations in natural and artificial
spin ice systems {[}3-6{]}. Excitations appear in natural systems
that are equivalent to the magnetic monopoles connected by Dirac strings,
which can be observed indirectly {[}1, 7-9{]}. In artificial systems
a pattern of magnetic nanoislands is created with lithographic techniques
(with a square lattice or kagome lattice) equivalent to natural spin
ice systems {[}10{]}. In the case of artificial spin ice systems,
the nanoisland are arranged such that there is frustration to minimize
dipolar interaction. This allows for designing geometries where emergent
excitations are present, equivalent to the magnetic monopoles of natural
spin ice systems. As well, with the artificial spin ice systems it
is possible to visualize the configuration of magnetic moments directly
with different microscopic techniques. This allows for their study
at ambient temperature and the simulation of the behavior of natural
spin ice systems {[}10-16{]}. Mengotti et al {[}17{]} recently published
the results of direct observations of emergent monopoles and associated
Dirac strings. These results are the first direct confirmation of
the reduction in the dimensionality of the system as a result of the
frustration in the artificial spin ice system. Mengotti\textquoteright{}s
work provides experimental evidence of reverse magnetization through
the Dirac string avalanches that join monopole-antimonopole pairs.
The same work provides results of Monte Carlo simulations that are
in agreement with experimental results. The excellent work of Mengotti
et al {[}17{]} provides very valuable information about the dynamic
of these elemental excitations and about nucleation and avalanches
in reverse magnetization. Equally, many questions were raised that
have lead to studying some aspects of the studied system by the authors
of this work {[}17{]}. For example, it is expected that the formation
of monopole-antimonopole pairs begins at the ends of the sample and
that the strings advance toward the center. In the aforementioned
report {[}17{]}, only the central region of the array is shown, without
the possibility of verifying the initial formation of the strings.
Equally, the Monte Carlo simulations of these and other authors, considering
infinite systems, are not very efficient when studying at the beginning
of the reversion of magnetization. One the central points of this
paper is to introduce the use of cellular automata to the field of
frustrated magnets. In this work, we study the effect of the aspect
ratio and size of the system on the formation of Dirac strings. A
study of the effect of disorder in the arrangement of nanomagnets
is incorporated in this work. We considered the system as a finite
array of magnetic nanoislands and took into account the possible disorder
in the value of the nanoisland moments. The simulation is carried
out using a determinist type of cellular automaton {[}18{]}, which
considers the coordination number of the elements of the grid. The
interaction among the nanoislands is studied under the magnetic charge
model and dipolar model.

\section{Magnetic nanoisland arrays and emergent monopoles}

The system studied in this work is a magnetic nanoisland array in
a hexagonal lattice. Figure 1 shows a scheme of the magnetic bars
arranged on the sides of a hexagon. Three nanomagnets converge in
the vertices of the hexagons, as can be appreciated in Figure 1. Geometrically,
we can define 2 non-equivalent vertices in the hexagonal lattice,
vertices A and B, respectively. The two non-equivalent vertices form
the unit cell of the entire array. The cell is shown in red in Figure
1. In the magnetic charge model, the charges are concentrated in these
vertices. We define the magnetic charge in a vertex as -1, when two
poles south and a north pole of the three nanomagnets that form the
vertex converge. Equally, we define the magnetic charge as +1, when
two poles north and a south pole converge in the considered vertex.
Figure 2 shows the array with an applied magnetic field. The upper
left of Figure 2 shows the array submitted to a field in the direction
of the negative axis $x$. In the lower left of the Figure the field
is shown directed to the right. The nanomagnets with the $x$ component
of the magnetic moment directed to the left are represented by gray
squares, and the nanomagnets with the $x$ component from the magnetic
moment directed to the right are represented by dark blue squares.
When the sample is totally magnetized in directions $+x$ or $-x$,
all the type A vertices have +1 (-1) charges, while all the B vertices
have -1 (+1) charges. This can be seen in Figure 2.

\begin{figure}[h]

\includegraphics[scale=0.4]{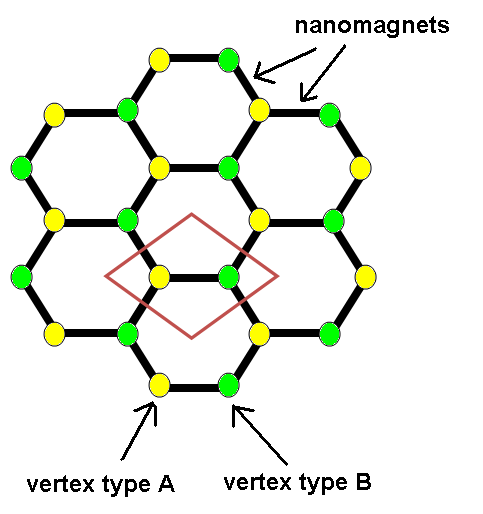}

Figure 1: Scheme of the nanomagnet array 
\end{figure}

\begin{figure}[h]
\includegraphics[scale=0.4]{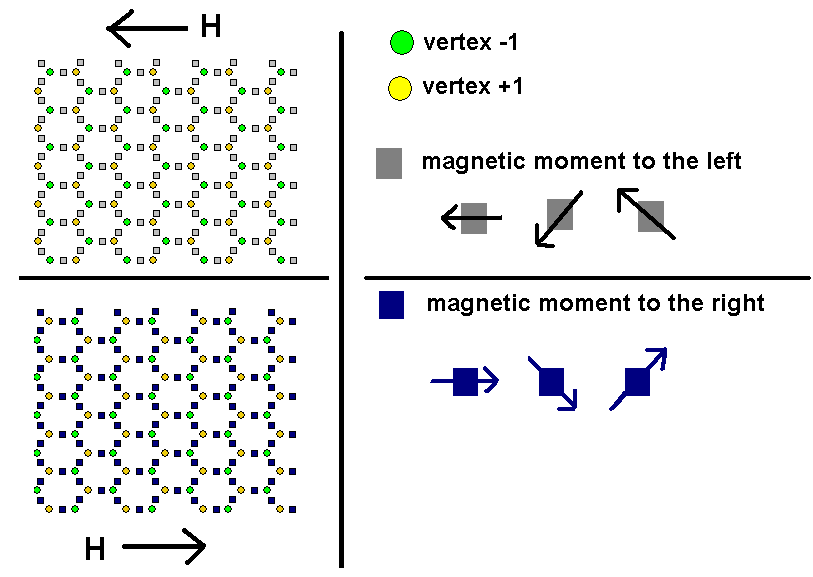}

Figure 2: Scheme of the nanomagnet array when the sample is totally
saturated by the influence of an external magnetic field.

\end{figure}

We suppose that we are under the condition of total magnetization,
with the magnetic field directed to the left (upper left of Figure
2). Under this condition, we define a positive and mobile monopole,
if a nanomagnet converging in a class A vertex inverts its magnetic
moment. The charge of vertex A goes from $q_{A}=-1\rightarrow q_{A}^{*}=+1\Longrightarrow\triangle q_{A}=+2$.
If this is produced in a B type vertex, we define a negative monopole
and would have $q_{B}=+1\rightarrow q_{B}^{*}=-1\Longrightarrow\triangle q_{B}=-2$,
where $q_{A}^{*}$ and $q_{B}^{*}$ represent the charge of the vertex
after the inversion and $q_{A}$ and $q_{B}$ represent the charges
of the vertices A and B, respectively, in the their initial states.
In this manner, when a nanomagnet inverts its magnetic moment, emerges
a monopole-antimonopole pair. Figure 3a - 3d, shows the generation
of a pair of mobile monopoles and their separation, giving rise to
a Dirac string. If the three nanomagnets that converge in a vertex
invert their moments, the condition $\triangle q_{A}=+2$ and $\triangle q_{B}=-2$,
is also generated, but in this case the monopoles remain trapped and
do not move through the sample (Figure 3e). 

\begin{figure}[H]
\includegraphics[scale=0.4]{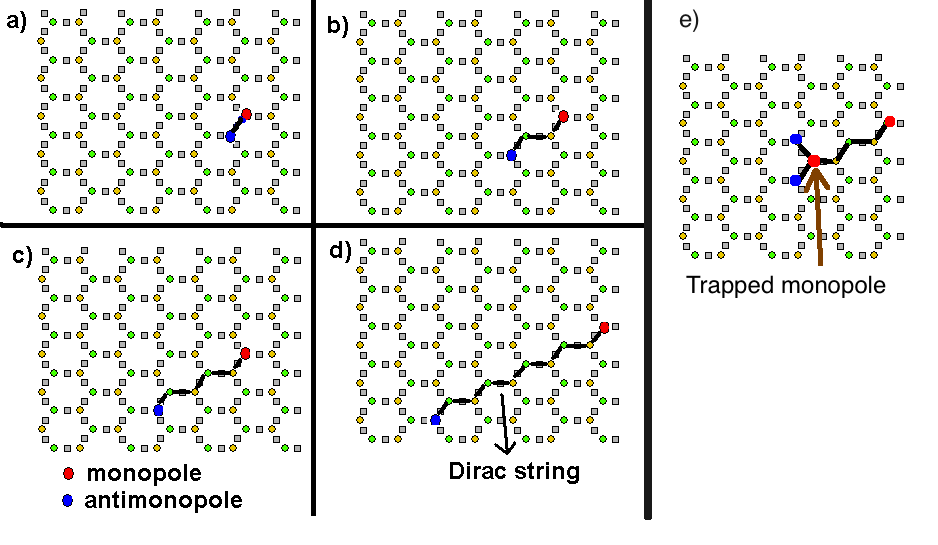}

Figure 3: Scheme of the creation of a monopole-antimonopole pair,
the associated Dirac string and trapped monopole.
\end{figure}

\section{Cellular automaton, the magnetic charge model and dipolar model.}

A cellular automaton (CA) is a mathematical structure used to model
the dynamics of complex systems. It is formed by many simple entities
that interact locally. A variety of models based on CA have been used
to efficiently study problems in biology, physics, chemistry, engineering
and material sciences {[}19-20{]}. They represent an excellent alternative
to models based on differential equations and to Monte Carlo algorithms
because they can simulate highly complex systems with a low computational
cost. The first attempt to use CA in the study of magnetism was the
model proposed by Vichniac {[}21{]}, which was subsequently developed
by Pomeau {[}22{]} and Hermann {[}23{]} and is termed the VPH model.
This is being used to resolve an Ising type spin system. To avoid
a \textquotedblleft{}feedback catastrophe\textquotedblright{}, the
automaton is updated in more than one step. The model functioned well
at high temperatures $\left(T>T_{C}\right)$, but failed at low temperatures.
Subsequently, Ottavi et al {[}24{]} used a microcanonical algorithm
in a CA to resolve the Ising spin system. A determinist version of
this model provided acceptable results at a low temperature {[}24{]}.
Owing to the popularity of the different types of Monte Carlo algorithms
used in problems associated with spins, the development of CA models
for these systems did not continue. In this work we used a CA model,
different from previous models, developed specifically to resolve
the dynamic of frustrated spins in artificial spin ice systems. This
model allows the simulation of spin ice systems efficiently {[}18{]}.

\subsection{Frustrated Cellular Automaton (FCA)}

This model was conceived for frustrated systems whose dynamic develops
at zero temperature or the equivalent. In the case of artificial ice
spin systems, each nanomagnet has a shape anisotropy with energy on
the order of $10^{4}\; K$. This means that the thermal fluctuations
in the configuration of moments are negligible at room temperature
and consequently the system behaves like a system at zero Kelvin.
To study this system in particular, we define the cells of the automaton
in the vertices of the hexagonal structure. The ends of three nanomagnets
converge in each vertex. Figure 4a shows the structure of the FCA.
The automaton is updated as follows: 

1. A class of diagonal nanomagnets (d1 or d2) is randomly selected
at each stage of the algorithm. 

2. It covers all the vertices of the automaton. 

3. The moment of the nanomagnet of the chosen class is inverted and
if the total energy decreases (and this difference is greater than
the energy barrier for investment), the change is accepted. The energy
term is explained in the next paragraph. This new configuration is
maintained in an auxiliary array. 

4. The same is then done with the horizontal nanomagnet (H), not considering
the nanomagnet of the non-selected class. 

6. The auxiliary configuration is copied in the definit configuration
and we return to step 1. 

In this way, we study the dynamic of emergent monopoles with a deterministic
model that allows studying finite systems, considering the effect
of the edge and the size of the system on the dynamic. As well, it
allows us to incorporate impurities in the sample and study the effect
of these impurities in the monopoles and associated Dirac strings. 

\begin{figure}[h]
\includegraphics[scale=0.4]{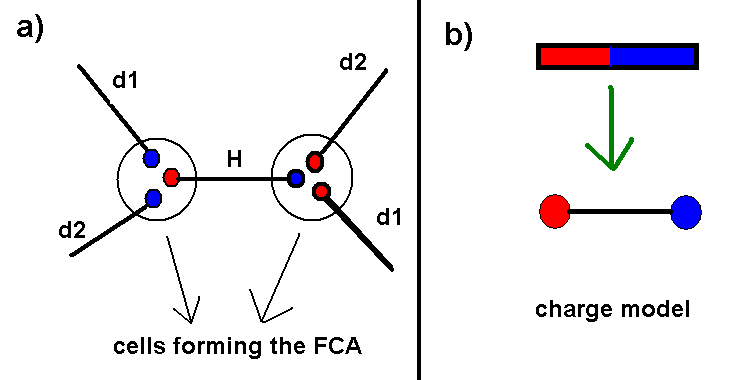}

Figure 4: a) Scheme of the FCA cells. b) Scheme of replacing the magnetic
moments by the magnetic charge model.
\end{figure}

\subsection{Magnetic charge model.}

The moment $\overrightarrow{m}$ of each nanoisland in this model
is replaced by two charges (one positive and the other negative),
located at the ends of the nanomagnet, as shown in Figure 4b. The
magnitude of each charge is $q=\frac{m}{l}$, where $l$ is the length
of the bar. The total charge in each vertex is the sum of the three
charges associated with the vertex. Vertex $j$ gives $Q_{j}={\displaystyle \sum_{k\epsilon j}q_{k}}$.
The total energy of the system is given by the expression:

\begin{equation}
U=\begin{cases}
\begin{array}{c}
\frac{1}{2}\frac{\mu_{0}}{4\pi}{\displaystyle \sum_{i,j}\frac{Q_{i}Q_{j}}{r_{ij}}},\: i\neq j\\
f_{i},\; i=j
\end{array}\end{cases}\label{eq:1}
\end{equation}

The term for $i\neq j$ takes into account the interaction among the
vertices of the array. The term $i=j$ considers the energy of the
site. This term considers the interaction among the ends of the three
nanomagnets that converge. Figure 5 shows a scheme of the configuration
of charges in each vertex and the parameters of associated length. 

\begin{figure}[h]
\includegraphics[scale=0.4]{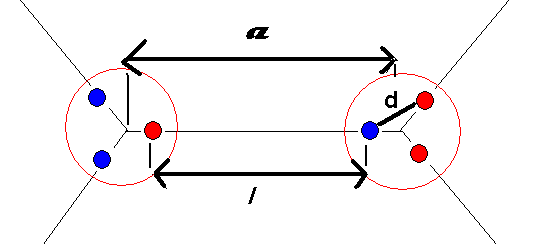}

Figure 5: Scheme of the configuration of charges in each vertex and
of the length parameters
\end{figure}

The energy in each vertex is given by the expression:

\begin{equation}
f_{i}=\frac{\mu_{0}}{4\pi}\left\{ \frac{q_{1}q_{2}}{d}+\frac{q_{1}q_{3}}{d}+\frac{q_{2}q_{3}}{d}\right\} \label{eq:2}
\end{equation}

In accordance with the parameters of the hexagonal lattice $d=\frac{\sqrt{3}}{2}\left(a-l\right)$.
Defining $q_{0}=\nicefrac{m}{l}$ and writing the energy in units
of $\frac{\mu_{0}q_{0}^{2}}{4\pi a}$, the total energy can be written
as:

\begin{equation}
U=\begin{cases}
\begin{array}{c}
\frac{1}{2}{\displaystyle \sum_{i,j}\frac{Q_{i}Q_{j}}{r_{ij}},\; i\neq j}\\
\frac{2}{\sqrt{3}\varepsilon}\left\{ q_{1}q_{2}+q_{1}q_{3}+q_{2}q_{3}\right\} ,\; i=j
\end{array}\end{cases}
\end{equation}

With $\varepsilon=1-\nicefrac{l}{a}$. When the automaton is updated,
the change in total energy is registered (using equation 3), that
is, under the magnetic charge model. The interaction of the charges
with the applied magnetic field is added to this term and the anisotropy
energy term .

\subsection{Dipolar interaction model}

The dipolar interaction between sites on the kagome lattice is given
by

\begin{equation}
U_{dip}=-\frac{\mu_{0}}{4\pi}\sum_{i,j(i\neq j)}\frac{1}{r_{ij}^{3}}\left\{ 3\left(\vec{m}_{i}\cdot\hat{r}_{ij}\right)\left(\vec{m}_{j}\cdot\hat{r}_{ij}\right)-\vec{m}_{i}\cdot\vec{m}_{j}\right\} 
\end{equation}
where $\vec{m}_{i}=\alpha_{i}m\widehat{e}_{i}$ is the magnetic moment
at site i. Here the pseudospin $\alpha_{i}=\pm1$ denotes the projection
of the spin onto the anisotropy (local ising) directions $\hat{e}_{i}$
(directed along the links of the honeycomb lattice) and $m$ is the
magnetic moment of a nanoisland. The interaction of the charges with
the applied magnetic field is added to this term and the anisotropy
energy term .

\section{Simulation and the results obtained}

\subsection{Monopoles and Dirac Strings with the charge model and dipolar model.}

The first system studied, is a sample of $\left(50\;\mu m\times50\;\mu m\right)$
and 5,800 nanomagnets, with impurities. The lattice constant (the
distance between two adjacent vertices) has a value of $a=577\: nm$.
The random magnetic moment of individual islands is given by $m=m_{0}\beta$,
where $\beta$ is a dimensionless Gaussian random variable with $\left\langle \beta\right\rangle =1$
and $s\equiv\left(\left\langle \left(\beta-\left\langle \beta\right\rangle ^{2}\right)\right\rangle \right)^{1/2}$.
In this first simulation, $s=0.13$ and the energy barrier for the
magnetic reversal is $98\times10^{4}kT$, with $T$, the room temperature.
This energy corresponds to a magnetic field of $\mu_{0}H=12,3$ $mT$
used in the simulation montecarlo {[}17{]}. The quantities $N_{T}$
and $\sigma_{M}$ shown in all the studies are represented in units
of $\frac{monopoles}{sites}$ and the magnetic field is shown as normalized
to the coercive field $H_{C}$. First, we do the simulation using
the dipolar model. Figure 6a shows the hysteresis curve, experimental
data of the hysteresis curve {[}17{]} and the total density of monopoles.
Figure 6b shows the density of mobile monopoles and experimental data
in function of the magnetic field. The agreement with experiment is
very good. The discrepancy in the size of the plateau is because in
this region, the simulation is very sensitive to the sample. This
sensitivity is also reflected in the Montecarlo simulation {[}17{]}.
It is important to note that at the left end the sample ends with
type ``A'' vertices and in the right end with type ``B'' vertices.
The monopoles generated in the latter vertices, when the moments of
the nanoislands are inverted are not mobile, and consequently cannot
migrant through the sample. The latter implies that the pair generated
in the ends only contributes a mobile monopole that is shifted toward
the center of the sample because of the applied magnetic field. Figure
7 shows a scheme of the complete sample (all the magnets) and the
region that is considered for the statistical analysis (red rectangle),
for values of the magnetic field near the coercivity. This figure
does not include the color of each vertex to better appreciate the
mobile monopoles. The impurities present in the sample (which were
deposited randomly) produce mobile monopoles pairs in every region
of the surface and both the positive and negative pole can move. This
is clearly illustrated in Figure 7, which simulates the dynamic of
system. We can note that some pairs of emergent monopoles appears
in the center of the sample. These emergent monopoles, which come
from impurities and are generated in the central part of the nanomagnet
arrays, move toward the ends, thus extending the Dirac string.

\begin{figure}[H]
\includegraphics[scale=0.3]{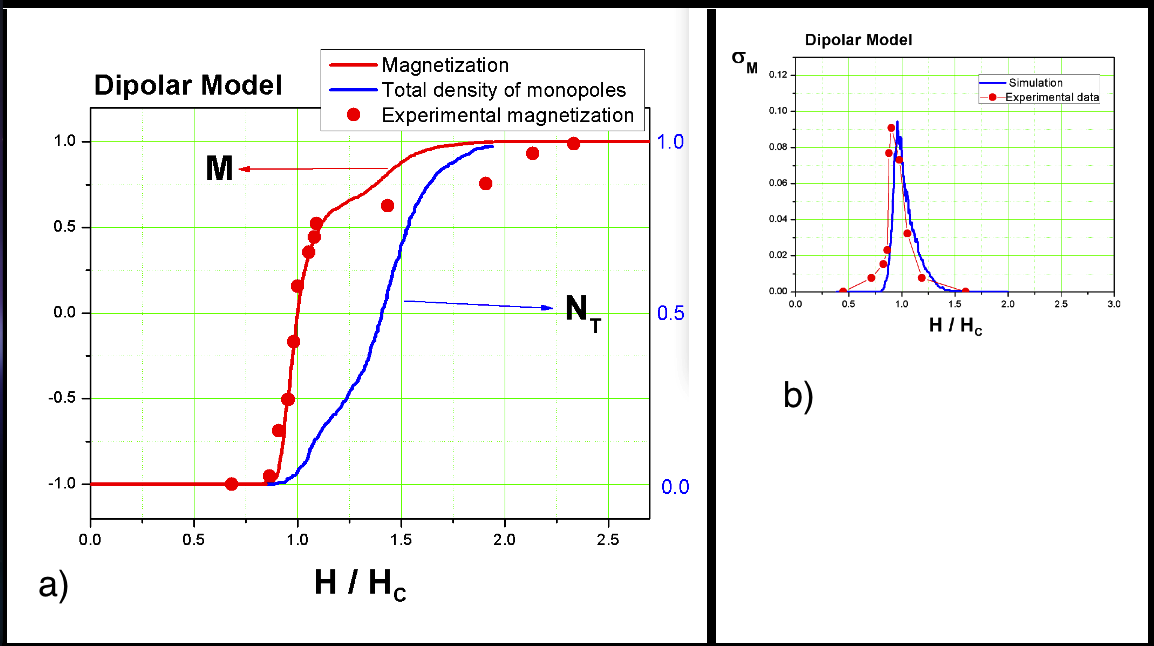}

Figure 6. Dipolar model $\left(\mu_{0}H_{c}=18.05\; mT\right)$. a)
Hysteresis curve (experiment {[}17{]} and simulation) and total density
of monopoles. b) Density of mobile monopoles (experiment {[}17{]}
and simulation).
\end{figure}

\begin{figure}[H]
\includegraphics[scale=0.45]{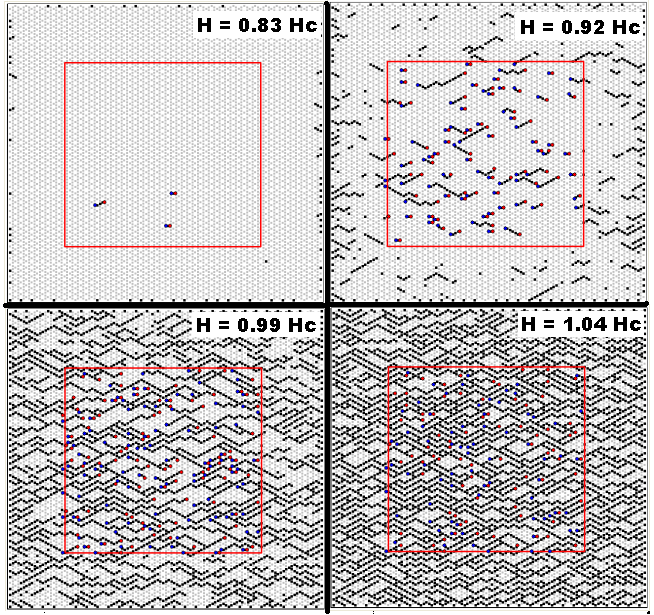}

Figure 7. Scheme of the simulation of the dynamic of monopoles and
Dirac strings for a system with 5,800 nanoislands with impurities
(dipolar model).
\end{figure}

We now examine the behavior of the system using the magnetic charge
model. All parameters are identical to the dipolar case, but now we
add the size of the magnets (required in this model). We use the value
$l=430\; nm$, which allows us to obtain, the experimental energy
of site. Figure 8a shows the hysteresis curve, experimental data of
the hysteresis curve {[}17{]} and the total density of monopoles.
Figure 8b shows the density of mobile monopoles and experimental data
in function of the magnetic field. We can appreciate some important
differences compared to dipolar simulation. First, the maximum density
of magnetic monopoles reaches a value of $0.082\;\frac{monopoles}{sites}$,
compared to the value of $0.095\;\frac{monopoles}{sites}$ in dipolar
simulation. 

Moreover, the total density of monopoles presents a curve with a slope
greater than in the case dipolar. The hysteresis curve has no plateau
characteristic of experimental data and dipolar simulation. These
differences in results are due to the energy of the site (charge model),
which prevents the initial reversal of nanomagnets H (Figure 4). This
behavior can be seen by comparing Figures 7 and 9.

We can appreciate a good agreement between the statistical results
of the dipolar simulation with experimental statistical results. However,
the images (XMCD) of the experimental paper in the first phase of
magnetic reversal, are more similar to the images obtained with the
simulation of magnetic charges.

\begin{figure}[H]
\includegraphics[scale=0.35]{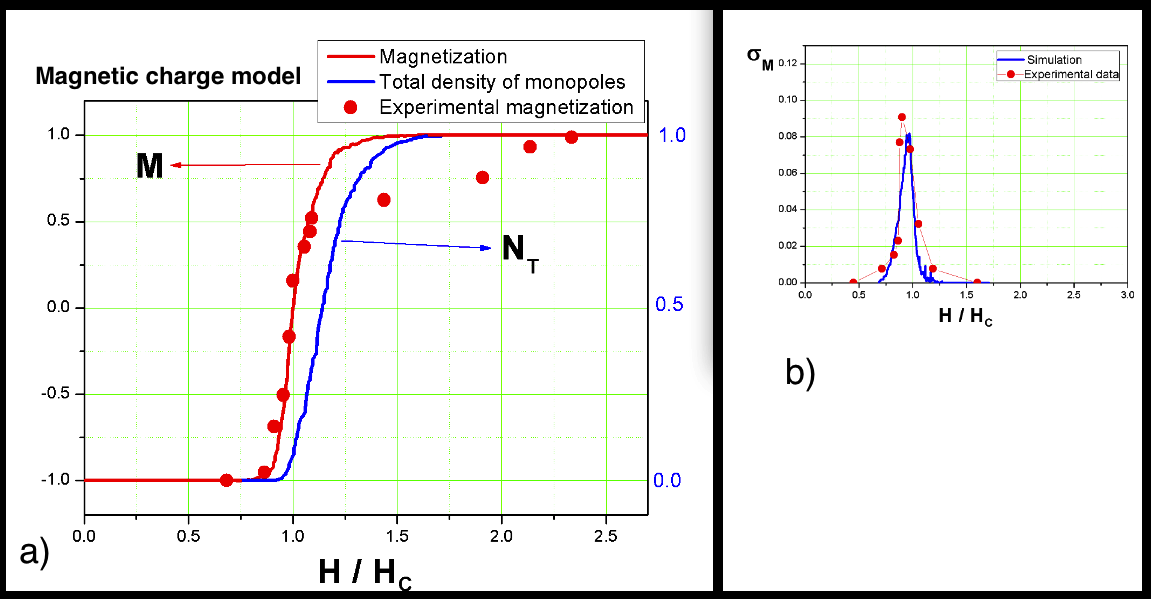}

Figure 8. Magnetic charge model $\left(\mu_{0}H_{c}=11.01\; mT\right)$.
a) Hysteresis curve (experiment {[}17{]} and simulation) and total
density of monopoles. b) Density of mobile monopoles (experiment {[}17{]}
and simulation).
\end{figure}

\begin{figure}[H]
\includegraphics[scale=0.45]{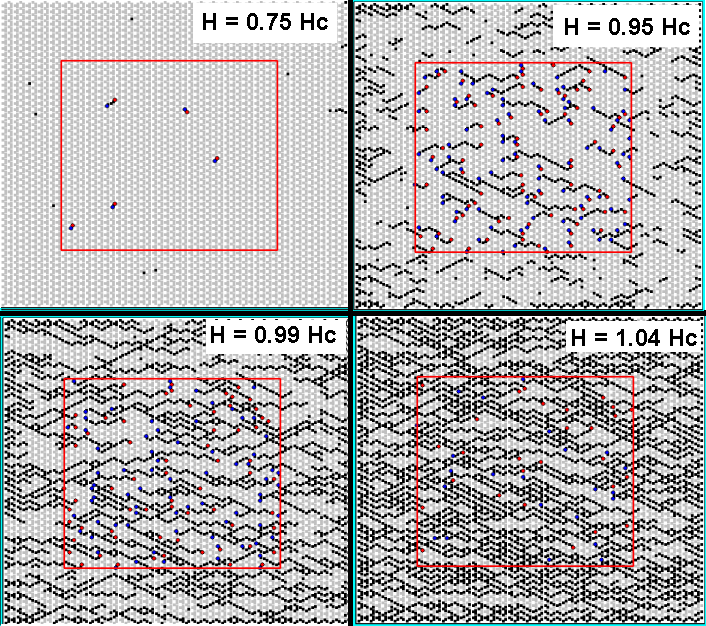}

Figure 9. Scheme of the simulation of the dynamic of monopoles and
Dirac strings for a system with 5,800 nanoislands with impurities
(charge model).

\end{figure}

\subsection{The effect of the size and aspect ration of the system on the dynamic
of emergent magnetic monopoles}

We summarize our study of the density of mobile monopoles in function
of the size of the system. For each size, the simulation was repeated
100 times. The error bars in the graphs correspond to standard error.
Figure 10 shows the maximum density of mobile monopoles, in units
of $\frac{monopoles}{sities}$, using the dipolar model, for two concentrations
of impurities $s=0.13$ and $s=0.25$. All systems, shown in Figure
10, have square geometry. Axis $x$ in Figure 10 shows the side of
the square in $\mu m$. From the Figure, we can appreciate that for
all systems, the density of the mobile monopoles is higher for the
case with $S=0.25$. As the size of the system decreases, the density
with impurities approaches to the maximum value of the $\frac{monopoles}{sites}$.
Our study verifies that the maximum density of magnetic monopoles,
decreases exponentially with system size. Figure 11 shows the same
studied, using charge model. We can verify a qualitative behavior
similar to the dipole model, but with lower density values. (discrepancy
explained above).

\begin{figure}[H]
\includegraphics[scale=0.3]{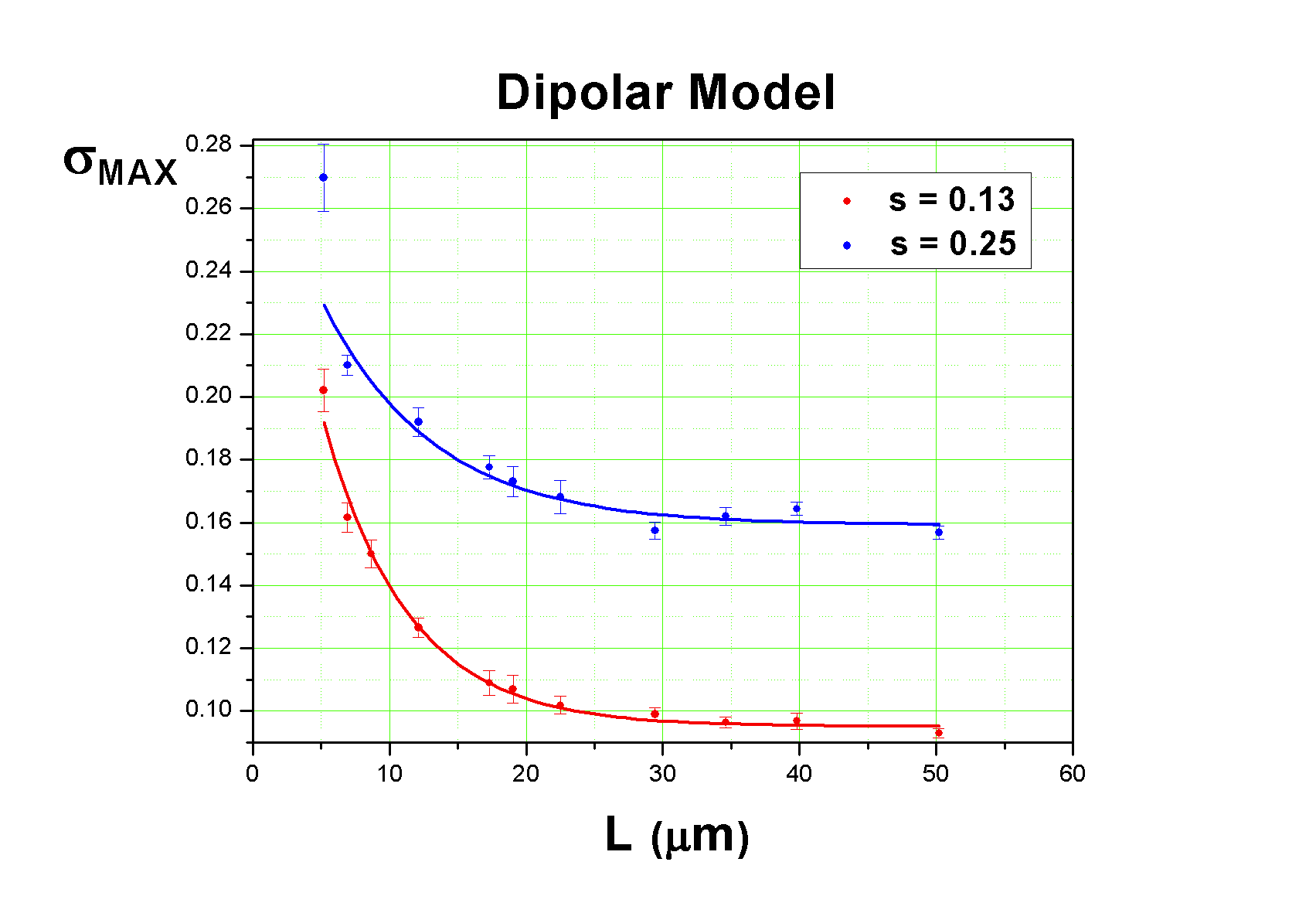}

Figura 10. Dependence of the density of the monopoles on the size
of the system. (Dipolar model).
\end{figure}

\begin{figure}[H]
\includegraphics[scale=0.3]{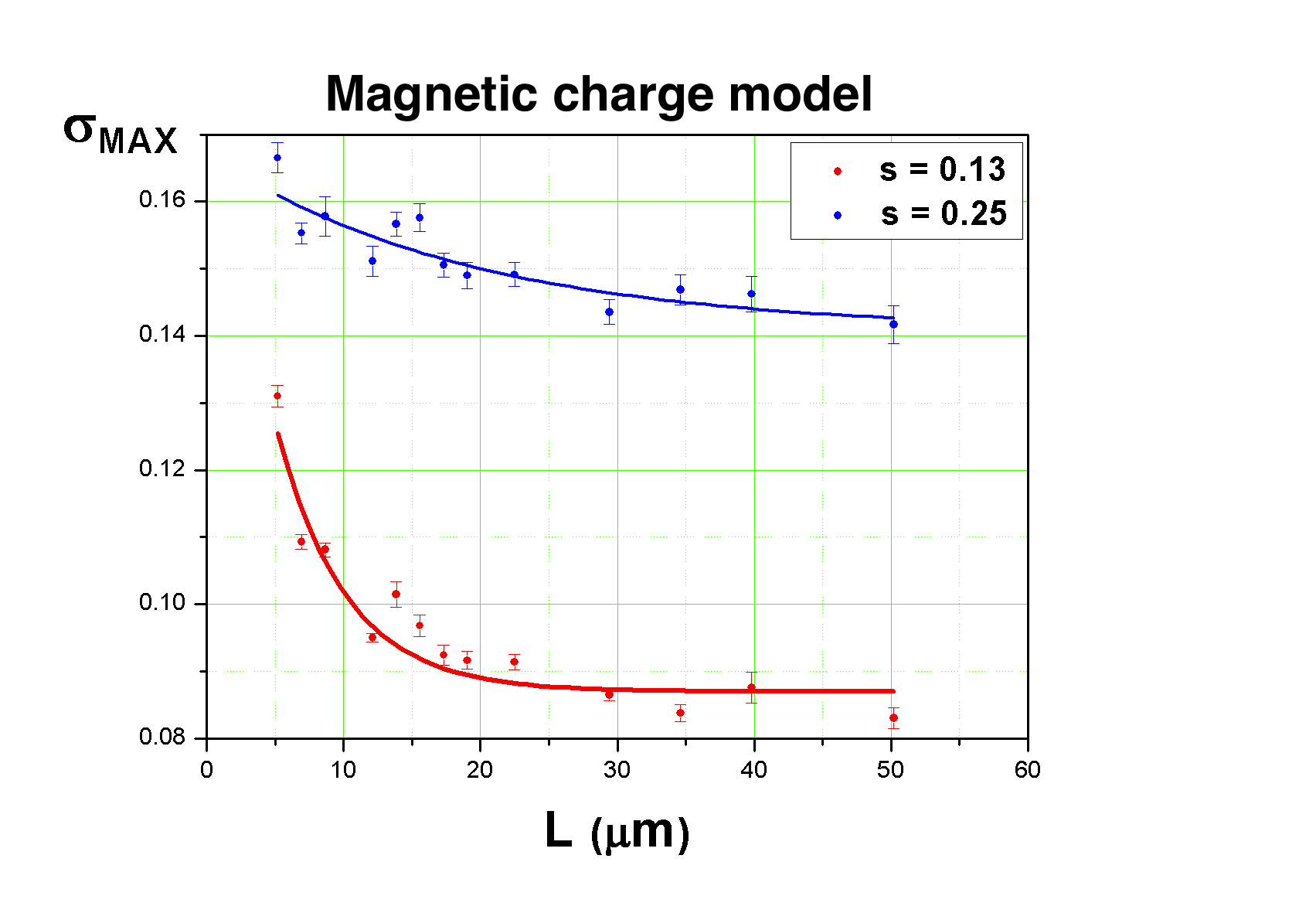}

Figure 11. Dependence of the density of the monopoles on the size
of the system. (Charge model).

\end{figure}

Finally, we show the results of the study considering the aspect ratio
of the system. Figure 12 shows the maximum value of the density of
mobile monopoles in function of the ratio $\frac{L_{X}}{L_{Y}}$,
where $L_{X}$ is the length of the sample in the direction parallel
to the magnetic field and $L_{Y}$ is the direction perpendicular
to the magnetic field. All systems studied have an impurity concentration,
equivalent to $s=0.13$. Our results show that the density of monopoles
is highly sensitive to the aspect ratio of the system.

\begin{figure}[H]
\includegraphics[scale=0.3]{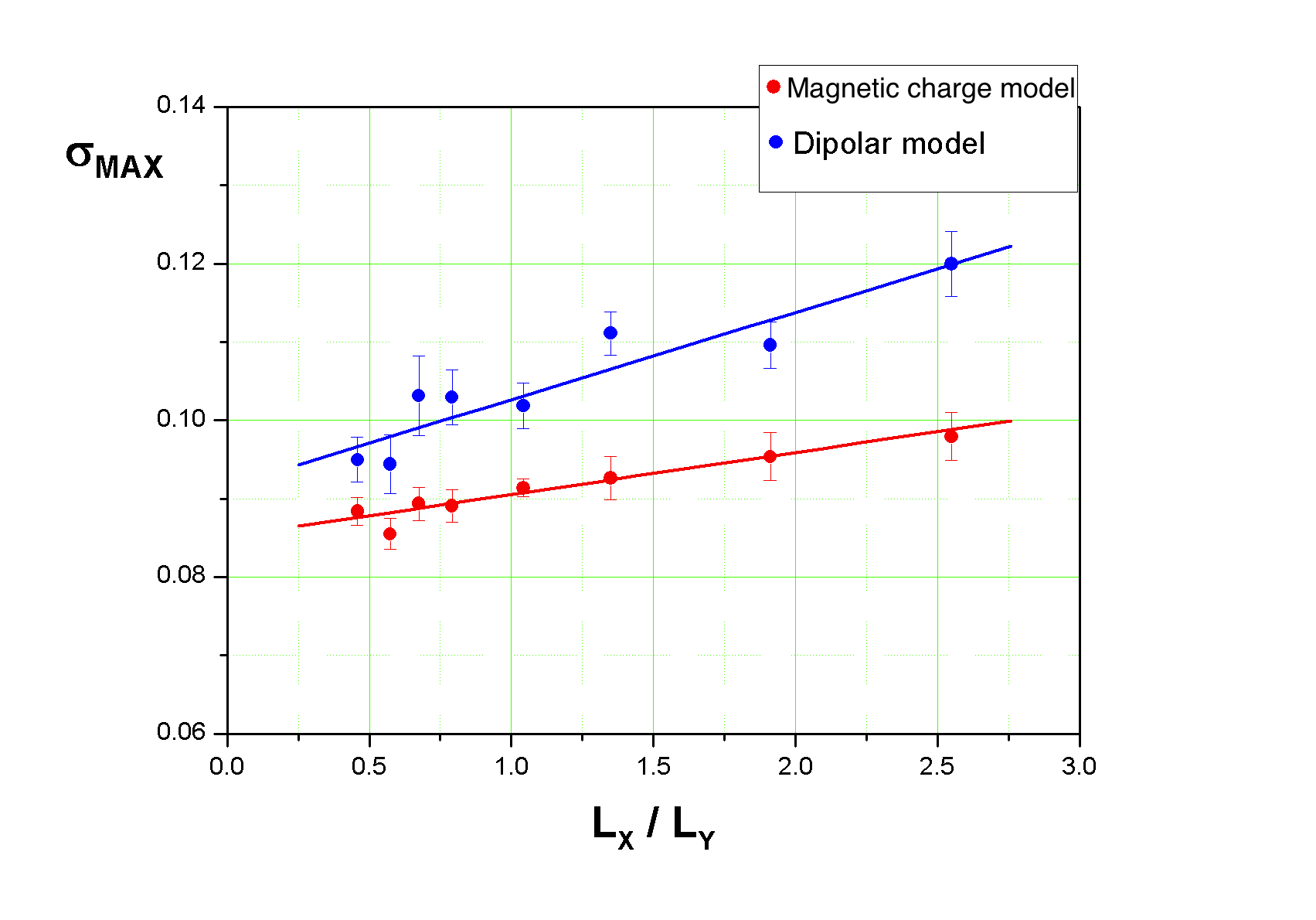}

Figure 12. Dependence of the density of the monopoles on the aspect
ratio of the system.
\end{figure}

\section{Conclusions}

In this work we have studied the dynamic of magnetic monopoles emergent
in an artificial spin ice system. An algorithm based on a frustrated
cellular automaton was used in the magnetic charge model and dipolar
model. The great advantage of the model is that it can make efficient
simulations of highly complex phenomena in real time with a minimum
of computational requirements. The model represent a perfect complement
to the methods based on Montecarlo algorithms, to study the elemental
physics of problems with classical and quantum entities. Our results
show that the number of emerging magnetic monopoles, depend of the
sample size, the aspect ratio and the concentration of impurities.
This allows us to consider a possible engineering in the creation
of these systems for applications, for example in information technologies
{[}25{]}.

Acknowledgments: The author acknowledge the financial support of FONDECYT
program grant 11100045.

\end{document}